\documentclass[smallextended]{svjour3}
\usepackage{amssymb}
\usepackage{amsmath}
\usepackage{amsfonts}
\usepackage{hyperref}
\usepackage[multiple]{footmisc}
\usepackage{xcolor}
\newcommand{\hoehe}[1]{\parbox[0pt][#1][c]{0cm}{}}
\begin{document}
\title{On causality in nonlinear vacuum electrodynamics of the Pleba\'nski class}

\author{Gerold O. Schellstede \and Volker Perlick \and Claus L\"{a}mmerzahl}
\institute{G. O. Schellstede, \email{gerold.schellstede@zarm.uni-bremen.de,\newline schellst@physik.fu-berlin.de} \at ZARM, University of Bremen, Am Fallturm, 28359 Bremen, Germany \and
V. Perlick, \email{volker.perlick@zarm.uni-bremen.de} \at ZARM, University of Bremen, Am Fallturm, 28359 Bremen, Germany \and C. L\"{a}mmerzahl, \email{claus.laemmerzahl@zarm.uni-bremen.de} \at ZARM, University of Bremen, Am Fallturm, 28359 Bremen, Germany and \at Institute of Physics, University of Oldenburg, 26111 Oldenburg, Germany}

\date{\today}

\maketitle

\begin{abstract}
We investigate the Pleba{\'n}ski class of electrodynamical
theories, i.e., theories of nonlinear vacuum electrodynamics that
derive from a Lorentz-invariant Lagrangian (or Hamiltonian). In 
any such theory the light rays are the lightlike geodesics of two 
optical metrics that depend on the electromagnetic background field.    
A set of necessary and sufficient conditions is found whose fulfillment 
secures that the optical metrics are causal in the sense that the light 
rays are lightlike or timelike with respect to the underlying space-time 
metric. Thereupon we derive conditions on the Lagrangian, or the Hamiltonian, 
of the theory such that the causality conditions are satisfied for all 
background fields. (The allowed values of the field strength tensor are
those for which the excitation tensor is finite and real.) The general 
results are illustrated with several examples. 
\end{abstract}

\section{Introduction}
In 1912 Gustav Mie made the first attempt to alter Maxwell's theory of 
vacuum electromagnetism in a way that leads to a consistent theory of point 
charges~\cite{Mie1,Mie2}. Despite the fact that Mie's theory is not 
gauge-invariant and therefore had to be abandoned, it strongly stimulated 
the search for nonlinear modifications of Maxwell's vacuum 
theory~\cite{Sommerfeld}. This led in particular to the nonlinear 
electrodynamical theory first published by Born in 1933~\cite{MB} and 
then, in a modified form, by Born and Infeld in 1934~\cite{BornInfeld1934}. 
Nowadays the Born-Infeld theory experiences a renaissance since 
Tseytlin~\cite{Tseytlin1999} pointed out that it can be derived as 
an effective theory from some string theories. 

After the development of quantum electrodynamics (QED) another remarkable 
attempt was made from a different direction in 1936 by Heisenberg and 
Euler~\cite{HeisenbergEuler1936} who derived a nonlinear effective 
classical theory of vacuum electrodynamics incorporating some effects 
from QED.

The Born, Born-Infeld and Heisenberg-Euler theories are examples of 
the so-called Pleba{\'n}ski class of nonlinear electrodynamics. 
This class comprises all theories that can be derived from a Lorentz-invariant
Lagrangian (or Hamiltonian). A comprehensive study of this class of
theories was brought forward by Pleba{\'n}ski~\cite{Plebanski1970};
important early contributions were also made by Boillat~\cite{Boillat}.

It is a characteristic feature of nonlinear electrodynamical
theories that the propagation of light is influenced by an electromagnetic
background field. In the geometric optics approximation light propagation
can be described in terms of rays. It was shown by Novello et 
al.~\cite{Novello} that for every theory of the Pleba{\'n}ski class 
the light rays are the lightlike geodesics of two optical metrics
which depend on the electromagnetic background field. The two optical 
metrics are associated with two different polarisation states. This 
result was rederived, using a different representation, by Obhukov and 
Rubilar~\cite{ObukhovRubilar2002} who also showed that the 
optical metrics are always of Lorentzian signature if non-degenerate.

As a consequence, we have three metrics of Lorentzian signature on
the space-time manifold when a theory of the Pleba{\'n}ski class and
an electromagnetic background field has been specified: The spacetime
metric which is assumed to be given from the outset, and the two 
optical metrics. This raises the question of how the light-cones of
these three metrics are related to each other. Important results 
in this direction have been found already by Goulart and Perez 
Bergliaffa~\cite{Costa} and in particular by Abalos et 
al.~\cite{Abalos}. In this paper we want to further elaborate 
on this question. In particular, it is our goal to give a 
necessary and sufficient condition for the optical metrics to be 
causal. Here and in the following we use the following terminology.
We say that the optical metrics are causal if all of their 
lightlike geodesics are timelike or lightlike with respect
to the space-time metric. If we adopt the usual interpretation of
(special or general) relativity according to which the light-cone
of the spacetime metric determines the maximal speed for signals, 
causality of the optical metrics is necessary to make sure
that light rays do not violate the speed limit for signals.
In this sense, nonlinear electrodynamical theories where at
least one of the two optical metrics violates the causality
condition have to be considered as unphysical. 

Without loss of generality, we will assume that the underlying
space-time metric is the Minkowski metric, i.e., we will 
restrict to special relativity. As all the conditions on the
light-cone structure are purely algebraic, the results immediatetly
carry over to general relativity: We just have to apply them 
to the light-cones on each tangent space.

The paper is organised as follows. In section 2 we consider 
a Lagrangian of the Pleba{\'n}ski class and we recall how the
characteristic equation (or, equivalently, the dispersion relation
or the Fresnel equation) can be written down if a background field has 
been specified. In sections 3 and 4 we decompose the characteristic
equation in terms of the two optical metrics and we derive inequalities
that guarantee causality of the optical metrics.
These inequalities are further evaluated in section 5. In section 6 
we derive conditions on the Lagrangian that guarantee causality of
the optical metrics \emph{for all} background fields. This question was
not touched upon in the above-mentioned articles~\cite{Costa,Abalos}.
In section 7 we rewrite these conditions in terms of a Hamiltonian, 
rather than a Lagrangian, formulation. In section 8 we illustrate 
our results with some examples.
 
\section{The characteristic equation}\label{AboutTheCharacteristicEquation}
We consider Minkowski spacetime in standard inertial coodinates, i.e., with
the (covariant) components of the metric tensor $\eta _{ik} = 
\mathrm{diag}(1,1,1,-1)$. We use Einstein's summation convention for Latin 
indices running from $1$ to $4$. 
Indices are lowered and raised with $\eta _{ik}$ and its inverse
$\eta ^{ik}$, respectively.

The electromagnetic theories we are interested in are characterised
by a Lagrangian $\mathcal{L}(F,G)$. Here 
\begin{equation}\label{eq:FG}
F=\frac{1}{2}\,F_{mn}F^{mn} 
\quad \text{and} \quad
G=-\frac{1}{4}\,F_{mn}\tilde F^{mn} 
\end{equation}
are the Lorentz invariants of the electromagnetic field strength
tensor $F_{ab}$; 
\begin{equation}\label{eq:tildeF}
\tilde F^{mn}\, = \, \frac{1}{2} \, 
\varepsilon^{mnab}F_{ab}
\end{equation}
is the Hodge dual of the field strength tensor, with 
$\varepsilon ^{mnab}$ denoting the totally antisymmetric 
Levi-Civita tensor, $\varepsilon
^{1234} = -1$. 

The universal Maxwell equations are
\begin{equation}\label{eq:Max1}
\partial_{[a} F_{bc]}=0 
\end{equation}
and
\begin{equation}\label{eq:Max2}
\partial_b H^{ab} \, = \, \frac{4\pi}{c}\,j^a\,,
\end{equation}
where a square bracket around indices means antisymetrisation.
The constitutive law that relates the excitation tensor $H^{ab}$ 
to the field strength tensor $F_{mn}$ is determined by the Lagrangian,
\begin{equation}\label{eq:const}
H^{ab}=-\frac{\partial \mathcal{L}}{\partial F_{ab}}
= -2\, \mathcal{L}_F \, F^{ab}+\mathcal{L}_G \,\tilde F^{ab} \, .
\end{equation}
Here and in the following we use the abbreviations
\begin{equation}\label{eq:LFLG}
\mathcal{L}_F = \dfrac{\partial \mathcal{L}}{\partial F} \, , \quad
\mathcal{L}_G = \dfrac{\partial \mathcal{L}}{\partial G} \, ,
\end{equation}
\begin{equation}\label{eq:LFLG2}
\mathcal{L}_{FF} = \dfrac{\partial ^2 \mathcal{L}}{\partial F ^2} \, , \quad
\mathcal{L}_{GG} = \dfrac{\partial ^2 \mathcal{L}}{\partial G ^2} \, , \quad
\mathcal{L}_{FG} = \dfrac{\partial ^2 \mathcal{L}}{\partial F \partial G} \, .
\end{equation}
It is the constitutive law (\ref{eq:const}) that 
distinguishes  different theories, while the Maxwell equations 
(\ref{eq:Max1}) and (\ref{eq:Max2}) are always the same. This class
of theories, with an arbitrary Lagrangian of the form $\mathcal{L}(F,G)$, 
is called the \emph{Pleba{\'n}ski class}. From (\ref{eq:const}) we
read that the standard Maxwell vacuum theory $H^{mn} = F^{mn}$ is
included with
\begin{equation}\label{eq:cMax}
\mathcal{L}(F) = - \dfrac{F}{2} \, .
\end{equation}

In this paper we will restrict to the geometric optics approximation. In
this approximation light propagation is determined by a \emph{characteristic
equation} (or \emph{eikonal equation}) which was derived, for an 
arbitrary theory of the Pleba{\'n}ski class, by Novello et al.~\cite{Novello}
and, in a different way, by Obukhov and 
Rubilar~\cite{ObukhovRubilar2002}. Here we use the same representation
as in Schellstede et al.~\cite{Michelson}. Then the characteristic 
equation reads
\begin{equation}\label{CharDiffGlg_FG}
\mathcal{L}_F\left\{M\eta^{ij}\eta^{kl}
+N\eta^{ij}F^{km}F^{l}_{\phantom lm}
+PF^{im}F^{j}_{\phantom jm}F^{kn}F^{l}_{\phantom ln}\right\}p_ip_jp_kp_l
=0\,
\end{equation}
with
\begin{gather}
\label{eq:M}
M:=\mathcal{L}_F^2+2\,\mathcal{L}_F \mathcal{L}_{FG}\,G
-\frac12\,\mathcal{L}_F \mathcal{L}_{GG}\,F+
\left(\mathcal{L}^2_{FG}-\mathcal{L}_{FF}\mathcal{L}_{GG}\right)G^2 \, ,
\\[0.1cm]
\label{eq:N}
N:=2\, \mathcal{L}_F \mathcal{L}_{FF}+\frac12\,\mathcal{L}_F \mathcal{L}_{GG}
+\left(\mathcal{L}^2_{FG}-\mathcal{L}_{FF}\mathcal{L}_{GG}\right)F \, ,
\\[0.1cm]
\label{eq:P}
P:=\mathcal{L}_{FF}\mathcal{L}_{GG}-\mathcal{L}^2_{FG}\,.
\end{gather}
The characteristic equation is to be viewed as a partial differential equation 
for a function $\psi (x^i)$. If $p_i  = \partial \psi/\partial x^i$ is a nowhere
vanishing solution of the eikonal equation, the function $\psi$ foliates
the space-time manifold into hypersurfaces $\psi = \mathrm{constant}$ which are
called \emph{characteristic surfaces} or \emph{wave surfaces}. If read as
an algebraic equation for the covector $p_i$, (\ref{CharDiffGlg_FG}) is
called the \emph{dispersion relation} or the \emph{Fresnel equation}.

\section{Causality of the optical metrics in the case 
$M\neq0$}\label{Kausalitaet_MNicht0}
In the case $M\neq0$ the characteristic equation is decomposable into 
a product of the form
\begin{equation}\label{eq:Q12}
\mathcal{L}_F \, Q_+ \, Q_-=0
\end{equation}
where
\begin{equation}\label{eq:QA}
Q_{\pm} = a^{ik}_{\pm} p_i p_k \,,
\end{equation}
with
\begin{equation}\label{eq:optmet}
a^{ik}_{\pm}=
\eta^{ik}+\sigma_{\pm} F^{il}F^{k}_{\phantom kl} \, ,
\end{equation}
\begin{equation}\label{eq:sigma}
\sigma _{\pm} = 
\dfrac{N}{2M} \pm \sqrt{\dfrac{N^2}{4M^2} - \dfrac{P}{M}}
\, .
\end{equation}
$a^{ik}_+$ and $a^{ik}_-$ are known as the \emph{optical metrics}. 
As
\begin{equation}\label{eq:sigmareal}
N^2-4MP= 
\Big( 2 \mathcal{L}_F \mathcal{L}_{FF} - 
\dfrac{1}{2} \mathcal{L}_F \mathcal{L}_{GG} - PF \Big) ^2
+ 4 \Big( \mathcal{L}_F \mathcal{L}_{FG}-PG \Big) ^2
\end{equation}
is a sum of two squares, the optical metrics are always real. Note
that we are free to multiply each of the optical metrics with a
non-zero factor, which may depend on the foot-point, without changing
the characteristic equation.

If 
\begin{equation}\label{eq:deta}
\text{Det}\left|a^{ik}_{\pm}\right|=
-\left(1+\sigma_{\pm} F-\sigma_{\pm} ^2G^2\right)^2
\end{equation}
does not vanish, we may introduce the contravariant components
$b^{\pm}_{ik}$ of the optical metrics via
\begin{equation}\label{eq:ab}
a^{ik}_{\pm} b^{\pm}_{kl} = \delta ^i_l \,  .
\end{equation}
Here we have to keep in mind that we agreed to raise and to lower
indices with the space-time metric, so in general $b^{\pm}_{ik}\neq 
a_{\pm ik}\equiv\eta_{im}\eta_{kn}a_{\pm}^{mn}$.
With the help of the well-known identities (see Plebanski~\cite{Plebanski1970})
\begin{equation}\label{InvariantenBeziehungen}
\tilde F{}_{mn}F^{nk}=G\,\delta^k_m\, , \quad
-F_{mn}F^{nk}+\tilde F{}_{mn}\tilde F{}^{nk}=F\,\delta^k_m\,,
\end{equation}
it is easy to verify that (\ref{eq:optmet}) can be rewritten as
\begin{equation}\label{eq:optmet2}
a^{ik}_{\pm}
=(1+\sigma_{\pm} F)\eta^{ik}+\sigma_{\pm} 
\tilde F{}^{il}\tilde F{}^{k}_{\phantom kl} 
\end{equation}
and that 
\begin{equation}\label{eq:b}
b^{\pm}_{ik}=
\frac{\left(1+\sigma_{\pm} F\right)\eta_{ik}
-\sigma_{\pm} F_{i}^{\phantom il}F_{kl}}{1+\sigma_{\pm}  F-\sigma_{\pm} ^2G^2}
=\frac{\eta_{ik}-\sigma_{\pm} \tilde F{}_{i}^{\phantom il}
\tilde F{}_{kl}}{1+\sigma_{\pm} F-\sigma_{\pm} ^2G^2}\,.
\end{equation}
From (\ref{eq:deta}) we read that the determinant of the optical
metric cannot be positive. As a consequence the signature
must be Lorentzian, i.e. $(+++-)$ or $(---+)$, if the 
determinant is non-zero. This result was already found 
by Obukhov and Rubilar~\cite{ObukhovRubilar2002}.

We first observe that we have to require $\mathcal{L}_F \neq 0$ because 
otherwise \emph{any} $p_i = \partial \psi/ 
\partial x^i$ satisfies the characteristic equation which cannot be
considered to be a reasonable law of light propagation. Then the 
characteristic equation (\ref{CharDiffGlg_FG}) splits into two 
equations
\begin{equation}\label{eq:HJ}
a_{\pm}^{ik}\, \dfrac{\partial \psi}{\partial x^i}
\, \dfrac{\partial \psi}{\partial x^k} = 0 \,, 
\end{equation}
with the optical metrics given by (\ref{eq:optmet}). Each of these two 
equations has the form of a Hamilton-Jacobi equation with the 
Hamiltonian $Q_{\pm}=a_{\pm}^{ik} p _i p _k $. The 
corresponding set of Hamilton equations 
\begin{gather}
\label{eq:Ham1}
\dot{x}{}^i=
\frac{\partial Q_{\pm}}{\partial p_i}=
2a^{ik}_{\pm} p_k=
2\eta^{ik}p_k-2\sigma_{\pm} F^{il}F^{\phantom lk}_{l}p_k
\\
\label{eq:Ham2}
\dot{p}{}_i=
-\frac{\partial Q_{\pm}}{\partial x^i}=
-p_mp_n \dfrac{\partial a^{mn}_{\pm}}{\partial x^i} 
\\
\label{eq:Ham3}
Q_{\pm} = a^{ik}_{\pm} p_i p_k = 0 
\end{gather}
determines the \emph{bicharacteristic curves} or \emph{rays}. For 
background material on the notion of bicharacteristic curves we refer 
to Courant and Hilbert \cite{CourantHilbert1962}. 

If the determinant
(\ref{eq:deta}) is non-zero, the covariant components (\ref{eq:b})
of the optical metrics are well defined and the Levi-Civita 
derivative $\nabla ^{\pm}$ of the optical metric $b^{\pm}_{ik}$ exists.
Then the Hamilton equations (\ref{eq:Ham1}), (\ref{eq:Ham2}) and 
(\ref{eq:Ham3}) require that the
rays are null geodesics of the optical metric, 
\begin{equation}\label{eq:rays}
\dot{x}{}^k \nabla ^{\pm}_k \dot{x}{}^i = 0 \, , \quad
p_i = \dfrac{1}{2} b^{\pm}_{ik} \dot{x}{}^k \, , \quad 
b^{\pm}_{ik} \dot{x}{}^i \dot{x}{}^k = 0 \, .
\end{equation}

For the rest of this section we fix one of the two optical metrics, $a_A^{ik}$, where the
index $A$ stands for $+$ or for $-$. We want to characterise the case that 
the optical metric $a_A^{ik}$ is non-degenerate and \emph{causal} in the 
sense that, for any solution to (\ref{eq:Ham1}) with (\ref{eq:Ham3}), 
the vector $\dot{x}{}^i$ satisfies $\eta _{ik}
\dot{x}{}^i \dot{x}{}^k \le 0$. The non-degeneracy condition
assures that there is a sphere's worth of directions tangent to
rays at each point, i.e., that it is possible to send a ray
in each spatial direction, and the causality condition assures 
that the rays are timelike or lightlike with respect to the 
space-time metric. The latter condition has to be fulfilled 
if we assume that the maximal speed for signals is determined 
by the null cone of the space-time metric. 

We know already that the non-degeneracy condition is 
satisfied if and only if $1+\sigma_AF-\sigma_A^2G^2 \neq 0$,
so we only have to investigate the causality condition.
If $\sigma _A = 0$, the optical metric
coincides with the space-time metric and the causality condition
is obviously satisfied, so it only remains to investigate the case
$\sigma _A \neq 0$. Moeover, if a covector $p_i$ is a principal null
covector of the electromagnetic field, i.e., if $F^{ji}p_i$ is a 
multiple of $\eta ^{ji}p_i$, the corresponding ray velocity $\dot{x}{}^i
= 2 a_A^{ik}p_k$ is lightlike with respect to the space-time metric, so
for these covectors the causality condition is never
violated. Therefore, what we have to investigate is the case that 
$1+\sigma_AF-\sigma_A^2G^2 \neq 0$ and $\sigma _A \neq 0$, and we
have to find a condition such that, for all solutions $p_i$ of 
(\ref{eq:Ham3}) that are not principal null covectors, the 
corresponding vector $\dot{x}{}^i = 2 a_A^{ik}p_k$ satisfies
$\eta _{ij} \dot{x}{}^i \dot{x}{}^j \le 0$.

To work this out, we fix such a covector $p_i$. Then the four
vectors 
\begin{equation}\label{eq:tetrad}
p^k = \eta ^{kl} p_l \, , \quad
\dot{x}{}^k= 2 a_A^{kl} p_l \, , \quad
\xi^k=F^{kl}p_l \, , \quad
\Xi^k=\tilde{F}{}^{kl}p_l-\sigma_AGF^{kl}p_l
\end{equation}
are linearly independent, which follows from the fact that 
$p_i$ is assumed not to be a principal null covector. With
the help of the identities (\ref{InvariantenBeziehungen}) 
one verifies that
\begin{equation}\label{BasisOrthonormal}
p_k\xi^k=p_k\Xi^k=\dot{x}_kp^k=\dot{x}_k\xi^k=\dot{x}_k\Xi^k=\xi^k\Xi_k=0
\end{equation}
and
\begin{equation}\label{SkalarprodukteBasisvekUnNorm}
\begin{split}
&p^ip_i=-\sigma_A\xi^i\xi_i\,;\quad\left(1+\sigma_AF-\sigma_A^2G^2\right)p^ip_i=-\sigma_A\Xi^i\Xi_i\,;\quad\\
&\frac{\dot{x}_k\dot{x}^k}{4}=-\left(1+\sigma_AF-\sigma_A^2G^2\right)p^ip_i\,.
\end{split}
\end{equation}
These relations fix the causal character of each of the four basis
vectors in dependence of the signs of $1+\sigma_AF-\sigma_A^2G^2$ and 
$\sigma _A$:
\begin{description}
\item[1)] \quad$1+\sigma_AF-\sigma_A^2G^2<0\,;\quad \sigma_A>0$

\hoehe{1cm}\qquad\qquad\qquad$\Rightarrow\quad p^ip_i >0\,;\quad 
\dot{x}_k\dot{x}^k> 0\,;\quad 
\xi^i\xi_i < 0\,;\quad 
\Xi^i\Xi_i > 0$

\item[2)] \quad$1+\sigma_AF-\sigma_A^2G^2<0\,;\quad \sigma_A<0$

\hoehe{1cm}\qquad\qquad\qquad$\Rightarrow\quad 
p^ip_i > 0\,;\quad 
\dot{x}_k\dot{x}^k > 0\,;\quad 
\xi^i\xi_i > 0\,;\quad 
\Xi^i\Xi_i < 0$

\item[3)] \quad$1+\sigma_AF-\sigma_A^2G^2>0\,;\quad \sigma_A>0$

\hoehe{1cm}\qquad\qquad\qquad$\Rightarrow\quad 
p^ip_i<0\,;\quad 
\dot{x}_k\dot{x}^k > 0\,;\quad 
\xi^i\xi_i > 0\,;\quad 
\Xi^i\Xi_i > 0$

\item[4)] \quad$1+\sigma_AF-\sigma_A^2G^2>0\,;\quad \sigma_A<0$

\hoehe{1cm}\qquad\qquad\qquad$\Rightarrow\quad 
p^ip_i > 0\,;\quad 
\dot{x}_k\dot{x}^k < 0\,;\quad 
\xi^i\xi_i > 0\,;\quad 
\Xi^i\Xi_i > 0$
\end{description}
We see that the causality condition is satisfied only in Case 4). 
In combination with our earlier observation that the causality 
condition is always satisfied if $\sigma _A =0$ we can summarise
the result of this section in the following way:

In the case $M \neq 0$ with $\mathcal{L}_F  \neq 0$ the optical metric
$a_A^{ik}$ is non-degegenerate and causal if and only if 
\begin{equation}\label{ok2}
1+\sigma_AF-\sigma^2_AG^2>0\,;\quad\sigma_A\leqq0\,.
\end{equation}
Note that for deriving this result we have not used that $\sigma _A$
is given by the specific expression (\ref{eq:sigma}), hence it is 
true for a metric of the form (\ref{eq:optmet}) with \emph{any}
$\sigma _A$.

Moreover, we mention that the causality condition (\ref{ok2}) can
also be derived with the help of the results of Abalos et 
al.~\cite{Abalos}. However, the derivation given here is simpler 
because it was not necessary to analyse the eigenvalue problem 
of the optical metric. 

\section{Causality of the optical metrics in the case $M=0$}\label{M0Fall}

In the case $M=0$ the characteristic equation 
(\ref{CharDiffGlg_FG}) again factorises,
\begin{equation}\label{eq:Q12ex}
\mathcal{L}_F \, Q_1 \, Q_2=0
\end{equation}
where
\begin{equation}\label{eq:QAex}
Q_{1/2} = a^{ik}_{1/2} p_i p_k \,, 
\end{equation}
but this time the optical metrics are
\begin{equation}\label{eq:omex}
a_1^{ij}= N\eta^{ij}+P F^{im}F^{j}_{\phantom jm} \, , \quad
a_2^{ij} = F^{im}F^{j}_{\phantom jm} \,.
\end{equation}
As in the case $M \neq 0$ we have to require $\mathcal{L}_F \neq 0$ because otherwise
the characteristic equation is satisfied identically. In analogy to the procedure
in the previous section we will investigate under what conditions the optical
metrics are non-degenerate and causal. 

In the following we make use of the fact that each optical metric is determined
uniquely only up to a non-zero factor which may depend on the foot-point.
If $N \neq 0$, we can divide the optical metric $a_1^{ij}$ by $N$. Then the
results of the previous section apply, with $\sigma _A$ replaced by $P/N$.
If $N =0$ and $P \neq 0$, we can divide $a_1^{ij}$ by P. Then the two
optical metrics coincide, so we only have to 
discuss $a_2^{ij}$ which we will do immediately. The case $N=P=0$ is
obviously forbidden by the non-degeneracy condition.

For the determinant of the metric $a_2^{ik}$ one gets
\begin{equation}\label{DetSpezialfall}
\text{Det}\left|a^{ik}_2\right|=-G^4\, ,
\end{equation}
so the non-degeneracy condition requires $G \neq 0$. For the Hamiltonian 
$Q_2 = a_2^{ik}p_ip_k$ Hamilton's equations read
\begin{gather}
\label{eq:Hamex1}
\dot{x}{}^i= \dfrac{\partial Q_2}{\partial p_i} =2 a_2^{ik}p_k =2 F^{im}F^k{}_mp_k \, ,
\\
\label{eq:Hamex2}
\dot{p}{}_i= \dfrac{\partial Q_2}{\partial x^i} = p_jp_k \dfrac{\partial a_2^{jk}}{\partial x^i}
= p_jp_k \dfrac{\partial (F^{jm}F^k{}_m)}{\partial x^i} \, ,
\\
\label{eq:Hamex3}
Q_2 = a_2^{ik}p_ip_k = F^{im}F^k{}_mp_ip_k = 0 \, .
\end{gather}
We will now prove that the causality condition $\eta _{ik} \dot{x}^i \dot{x}{}^k \le 0$ cannot
hold for all solutions of these equations if the non-degeneracy condition $G \neq 0$ is
satisfied.

To that end we choose a non-zero covector $p_i$ that satisfies (\ref{eq:Hamex3}) and is
not a principal null covector of the electromagnetic field, i.e.,  $\xi^k:=F^{kl}p_l$ and
$p^k$ are linearly independent. Such a covector exists because it is well known
that for an electromagnetic field with $G \neq 0$ there are precisely two linearly
independent principal null covectors, see e.g. the Appendix of Abalos et al.~\cite{Abalos}.
Then we find from (\ref{eq:Hamex3}) and from the antisymmetry of $F^{kl}$ that
\begin{equation}\label{eq:pxi}
\xi_k \xi ^k =0 \, , \quad \xi ^k p_k = 0 \, .
\end{equation}
As $\xi ^k$ and $p^k$ are linearly independent, these two conditions imply that $p^k$
is spacelike with respect to the space-time metric, $p_kp^k > 0$. On the other hand, we
find from (\ref{eq:Hamex1}) that
\begin{equation}\label{eq:pdx}
\dot{x}^k\dot{x}_k=4G^2p^ip_i\, ,
\end{equation}
so the non-degeneracy condition $G \neq 0$ yields $\dot{x}{}^k \dot{x}{}_k > 0$, i.e., 
the causality condition is necessarily violated.

We summarise the result of this section in the following way: In the case $M = 0$
with $\mathcal{L}_F \neq 0$ it is impossible that both optical metrics are non-degenerate and
causal.

Note that the results of this section are irrelevant for the case of 
$\mathcal{L}(F)$-theories, i.e., for Lagrangians that are independent
of $G$, because in this case the condition $M\neq0$ is equivalent to 
$\mathcal{L}_F \neq 0$.

\section{Evaluation of the causality conditions for both optical metrics}
\label{Abschnitt5}
In this short section we evaluate the causality condition (\ref{ok2}) for 
the case that it holds both for $\sigma _+$ and for $\sigma _-$. In view of 
$\sigma_{\pm}$ given by (\ref{eq:sigma}) we assume that $\sigma_+\geqq
\sigma_-$ but otherwise the specific form of $\sigma_{\pm}$ is not used
in this section. We assume $G\neq0$ and treat the case $G=0$ in the end 
as a limit.

For any given $F$ and $G \neq 0$, we define a real-valued function $T$ by
\begin{equation}
T(x):=x^2G^2-xF-1
\end{equation}
such that the condition $1+\sigma_AF-\sigma^2_AG^2>0$ is equivalent to $T(\sigma_A)<0$
Obviously, the roots of $T(x)=0$ are
\begin{equation}
x_{\pm}=\frac{F}{2G^2}\pm\sqrt{\frac{F^2}{4G^4}+\frac{1}{G^2}}
\end{equation}
with $x_+>0$ and $x_-<0$. Moreover, the only extremum (minimum) of $T(x)$ is given by
\begin{equation}\label{eq:Tp}
 T'(x_m)=2 x_m G^2-F=0\quad\Rightarrow\quad 
x_m=\frac{F}{2G^2}\quad\text{with}\quad
T''(x_m)= 2G^2\,.
\end{equation}
Therefore $T(x)<0$ together with $x\leqq0$ is equivalent to
\begin{equation}\label{KausalGNicht0}
0\geqq x > \frac{F}{2G^2}-\sqrt{\frac{F^2}{4G^4}+\frac{1}{G^2}}
\, .
\end{equation}
This is satisfied both by $x=\sigma_+$ and by $x= \sigma _-$ if and only if
\begin{equation}\label{eq:sigmain}
0\geqq\sigma_+ \, , \quad 
\sigma_->\frac{F}{2G^2}-\frac{1}{G^2}\sqrt{\frac{F^2}{4}+G^2}\,.
\end{equation}
These conditions are also true in the case $G=0$. For $G \to 0$, 
the second inequality is automatically satisfied ($\sigma _- > - \infty$) 
if $F \le 0$ and it requires $\sigma _- > -1/F$ if $F>0$. This can be
easily seen by Taylor expanding the square-root about $G=0$.

\section{Causality conditions on the Lagrangian}\label{sec:causLag}
The results of the preceding sections allow us to check if causality
holds if not only the the Lagrangian but also the background field has
been specified. In this section we want to derive conditions
\emph{on the Lagrangian} that guarantee causality for all allowed
background fields. Note that, in general, not all values of $F$ and $G$ are allowed. 
The best known counter-example is the Born-Infeld theory, 
see section~\ref{subsec:BI} below, where the Lagrangian involves a 
square-root and only those values of $F$ and $G$ are allowed for which 
the expression under the square-root is positive. In this section we 
will not specify the allowed values of $F$ and $G$ but we
will assume that they form a connected domain in the $(F,G)$-plane that
contains the point $(F,G) = (0,0)$. 

We have seen that a reasonable law of light propagation requires 
$\mathcal{L}_F \neq 0$. If this condition is supposed to hold for all 
allowed values of $F$ and $G$, continuity implies that either 
$\mathcal{L}_F <0$ or $\mathcal{L}_F>0$ on the entire domain of
allowed values. For the following discussion we assume that
\begin{equation}\label{eq:LFneg}
\mathcal{L}_F<0 
\end{equation}  
because this includes the standard Maxwell vacuum case which, with our
conventions, is given by (\ref{eq:cMax}). We will then get the
corresponding results for the case $\mathcal{L}_F >0$ easily at the end 
of the section by replacing $\mathcal{L}$ with $-\mathcal{L}$. Note 
that in principle we are free to multiply the Lagrangian with a non-zero 
constant factor, positive or negative, without changing the 
characteristic equation. We just have to keep in mind that then we also 
have to multiply the right-hand side of the constitutive law 
(\ref{eq:const}) with the same factor. 

To investigate the causality conditions on the Lagrangian we first 
recall from the preceding sections that causality of both optical metrics
is possible only if $M\neq 0$. As $M = \mathcal{L}_F^2>0$ for $(F,G)=(0,0)$, 
continuity requires
\begin{equation}\label{eq:Mpos}  
M>0
\end{equation}
on the entire domain of allowed values. 

In combination with (\ref{eq:LFneg}) and (\ref{eq:Mpos}) we have to evaluate the
two inequalities (\ref{eq:sigmain}). With $\sigma _+$ given by (\ref{eq:sigma}), the
first inequality requires 
\begin{equation}\label{eq:sigmapin}
\dfrac{N}{2M} + \sqrt{ \dfrac{N^2}{4M^2} - \dfrac{P}{M}} \le 0 \, .
\end{equation}
As $M>0$, this is true if and only if
\begin{equation}\label{eq:PNin}
N \le 0 \, , \quad P \ge 0 \, .
\end{equation}  
The second inequality in  (\ref{eq:sigmain}) requires 
\begin{equation}\label{eq:sigmapin1}
\dfrac{N}{2M} - \sqrt{\dfrac{N^2}{4M2}- \dfrac{P}{M}} >
\dfrac{F}{2G^2} - \sqrt{\dfrac{F^2}{4G^4}+\dfrac{1}{G^2}} \, .
\end{equation}
We evaluate this condition first for the case that $P=0$ and then for
the case that $P \neq 0$. If $P=0$, (\ref{eq:sigmapin1}) simplifies to
\begin{equation}\label{eq:sigmapin2}
\dfrac{N}{M}  >
\dfrac{F}{2G^2} - \sqrt{\dfrac{F^2}{4G^4}+\dfrac{1}{G^2}} 
\end{equation}
where we have used that $N \le 0$.  Multiplication with the 
strictly positive factor $(F+ \sqrt{F^2+4G^2})/(2G^2)$ results in
\begin{equation}\label{eq:M0}
M> - \dfrac{N}{2} \Big( F + \sqrt{F^2+4G^2} \Big) \, .
\end{equation}
Note that (\ref{eq:M0}) implies $M>0$, so the latter condition
need not be stated separately. 

In the case $P \neq 0$, (\ref{eq:PNin}) requires $P>0$.  Multiplication
of (\ref{eq:sigmapin1}) with the strictly positive factor $(-N-\sqrt{N^2
-4MP}) (F+ \sqrt{F^2+4G^2}) /(4MG^2)$ results in
\begin{equation}\label{eq:Mg}
-N - P\left(F+\sqrt{F^2+4G^2}\right) > \sqrt{N^2-4MP} \, .
\end{equation}
This inequality is equivalent to the conditions that the left-hand
side is positive and that the inequality is true with both sides
squared. These two conditions read
\begin{equation}\label{eq:NMg1}
N < -P \Big( F + \sqrt{F^2+4G^2} \Big) \, ,
\end{equation}
\begin{equation}\label{eq:NMg2}
M > -PG^2- \dfrac{1}{2} \Big(N+PF \Big) 
\Big( F + \sqrt{F^2+4G^2} \Big)
\, .
\end{equation}
(\ref{eq:NMg1}) implies $N<0$, and both conditions together
imply $M>0$. We can summarise the results in the following way.
For a Lagrangian with $\mathcal{L}_F < 0$ causality holds if and only
if for all allowed background fields
\begin{equation}\label{eq:causcon1}
P \ge 0 \, ,
\end{equation}
\begin{equation}\label{eq:causcon2}
N \le -P \big( F + \sqrt{F^2+4G^2} \big) \, ,
\end{equation}
\begin{equation}\label{eq:causcon3}
M > -PG^2- \dfrac{1}{2} \Big( N+PF \Big) 
\Big( F + \sqrt{F^2+4G^2} \Big)
\, ,
\end{equation}
where the equality sign in (\ref{eq:causcon2}) can hold
only if $P=0$. 

Inserting the definitions of $M$, $N$ and $P$ into these expressions
results in
\begin{equation}\label{eq:causcon4}
\mathcal{L}_{FF}\mathcal{L}_{GG} \ge \mathcal{L}_{FG}^2 \, ,
\end{equation}
\begin{equation}\label{eq:causcon5}
2\mathcal{L}_F \mathcal{L}_{FF}+ \dfrac{1}{2} \mathcal{L}_F \mathcal{L}_{GG} \le - 
\Big( \mathcal{L}_{FF} \mathcal{L}_{GG}- \mathcal{L}_{FG}^2 \Big) \sqrt{F^2+4G^2} \, ,
\end{equation}
\begin{equation}\label{eq:causcon6}
\begin{split}
\mathcal{L}_F<-F \mathcal{L}_{FF}&+\frac14\,F \mathcal{L}_{GG}-2 G \mathcal{L}_{\text{FG}}\\
&- \left( \mathcal{L}_{FF}+ \frac{1}{4}\,\mathcal{L}_{GG}\right)
\sqrt{F^2+4 G^2}=:\alpha
\end{split}
\end{equation}
where equality in (\ref{eq:causcon5}) can hold only if
$\mathcal{L}_{FF} \mathcal{L}_{GG} = \mathcal{L}_{FG}^2$. 
For theories with $\mathcal{L}{}_F < 0$, the conditions (\ref{eq:causcon4}),
(\ref{eq:causcon5}) and (\ref{eq:causcon6}) are equivalent to 
\begin{equation}\label{eq:causcon7}
 \mathcal{L}_{FF}\mathcal{L}_{GG} \geq \mathcal{L}_{FG}^2\,,
\end{equation}
\begin{equation}\label{eq:causcon8}
\mathcal{L}_{FF} \geq 0 \, , \quad \mathcal{L}_{GG} \geq 0 \, , 
\end{equation}
\begin{equation}\label{eq:causcon9}
\mathcal{L}_F<-F \mathcal{L}_{FF}+\frac14\,F \mathcal{L}_{GG}-2 G \mathcal{L}_{\text{FG}}-\left( \mathcal{L}_{FF}+ \frac{1}{4}\,\mathcal{L}_{GG}\right)
\sqrt{F^2+4 G^2}\,.
\end{equation}
To prove this, it is easy to see that (\ref{eq:causcon4}) and (\ref{eq:causcon5}) imply
(\ref{eq:causcon8}). Conversely, one can verify with a bit of algebra that  (\ref{eq:causcon7})
and (\ref{eq:causcon8}) imply $\alpha \le 0$ where $\alpha$ is defined in (\ref{eq:causcon6}); 
with this information  at hand, it can be shown that (\ref{eq:causcon7}), (\ref{eq:causcon8})
and (\ref{eq:causcon9}) imply (\ref{eq:causcon5}). 

We have derived the causality conditions here for Lagrangians
with $\mathcal{L}_F < 0$. The results for the case $\mathcal{L}_F>0$ follow
immediately by replacing $\mathcal{L}$ with $-\mathcal{L}$. Then (\ref{eq:causcon7})
remains unchanged whereas the inequality signs in (\ref{eq:causcon8}) and 
 (\ref{eq:causcon9}) have to be reversed.

For an $\mathcal{L}(F)$-theory, i.e., if the Lagrangian is independent
of $G$, the causality conditions (\ref{eq:causcon7}), (\ref{eq:causcon8}) and
(\ref{eq:causcon9})drastically simplify to
\begin{equation}\label{eq:causconF1}
\mathcal{L}_{FF} \ge 0 \, ,
\end{equation}
\begin{equation}\label{eq:causconF2}
-\mathcal{L}_F - \mathcal{L}_{FF}F > \mathcal{L}_{FF} \sqrt{F^2+4G^2} \, .
\end{equation}
The striking feature of (\ref{eq:causconF2}) is that the left-hand side is 
independent of $G$. If $\mathcal{L}_{FF} \neq 0$, the right-hand side increases 
with $G$, so for any allowed $F$ we can find a sufficiently big value of 
$G$ such that (\ref{eq:causconF2}) is violated. Here it is crucial that all 
values of $G$ are allowed because a Lagrangian of the form $\mathcal{L}(F)$ 
and therefore the resulting field equations can 
restrict only the values of $F$ but not those of $G$. In other words, 
\emph{any} $\mathcal{L}(F)$-theory violates the causality condition for some allowed 
field values unless $\mathcal{L}_{FF}$ is identically zero. The latter case is just 
the standard Maxwell vacuum theory.

\section{Causality conditions in the Hamiltonian formulation}\label{sec:causHam}
As an alternative to the Lagrangian formalism we have used so far, one can introduce
a Hamiltonian formalism. The basic equations are well-known since the pioneering
work of Pleba{\'n}ski~\cite{Plebanski1970}. The passage from the Lagrangian to 
the Hamiltonian formulation is possible if the constitutive law (\ref{eq:const})
can be solved for the field strength tensor $F_{ab}$.

The Hamiltonian is then derived from the Lagrangian via a Legendre transformation
\begin{equation}\label{eq:Hamilton}
\mathcal{H}(H^{ab})=-\frac{1}{2}H^{mn}F_{mn}-\mathcal{L}(F_{ab})
\end{equation}
where on the right-hand side the field strength tensor has to be expressed 
in terms of the excitation. For a theory of the Pleba\'nski class the 
Hamiltonian is a function of the two excitation invariants
\begin{equation}\label{eq:RS}
R=-\frac{1}{2} \,H^{ab}H_{ab}
\quad \text{and} \quad
S=\frac{1}{4}H_{ab}\tilde H{}^{ab} \, .
\end{equation}
In the Hamiltonian formalism the constitutive law 
reads 
\begin{equation}\label{eq:constH}
F_{ab}=- \dfrac{\partial \mathcal{H}}{\partial H^{ab}} =
2\,\mathcal{H}_R\,H_{ab}-\mathcal{H}_S\,\tilde H_{ab} \, .
\end{equation}
We have used already in~\cite{Michelson} the fact that all equations of 
the Hamiltonian formalism can be obtained from the corresponding equations 
in the Lagrangian formalism by the replacement rules
\begin{equation}\label{eq:duality}
F^{mn} \hookrightarrow \tilde H{}^{mn} \, , \quad
\mathcal{L} \hookrightarrow \mathcal{H} \, .
\end{equation}
As $F^{mn} \hookrightarrow \tilde H{}^{mn}$ implies
\begin{equation}\label{DualRotTab}
\begin{split}
\tilde F{}^{mn}\hookrightarrow -H{}^{mn} \, , \quad
F\hookrightarrow R \, , \quad G\hookrightarrow S\ \, ,
\end{split}
\end{equation}
the characteristic equation can be rewritten in terms of the 
Hamiltonian as
\begin{equation}\label{CharDiffGlg_RS}
\mathcal{H}_R\left\{\hat M\eta^{ij}\eta^{kl}+\hat N\eta^{ij}\tilde H{}^{km}\tilde H{}^{l}_{\phantom lm}+\hat P\tilde H{}^{im}\tilde H{}^{j}_{\phantom jm}\tilde H{}^{kn}\tilde H{}^{l}_{\phantom ln}\right\}p_ip_jp_kp_l=0\,
\end{equation}
with 
\begin{equation}
\begin{split}
&\hat M=\mathcal{H}_R^2+2\,\mathcal{H} _R \mathcal{H}_{RS}\,S-\frac12\,\mathcal{H}_R \mathcal{H}_{SS}\,R +\left(\mathcal{H}^2_{RS}-\mathcal{H}_{RR} \mathcal{H}_{SS}\right)S^2 \,,\\
&\hat N=2\,\mathcal{H}_R\mathcal{H}_{RR}+\frac12\,\mathcal{H}_R \mathcal{H}_{SS} +\left(\mathcal{H}^2_{RS}-\mathcal{H}_{RR}\mathcal{H}_{SS}\right)R \,,\\
&\hat P=\mathcal{H}_{RR}\mathcal{H}_{SS}-\mathcal{H}^2_{RS}\,.
\end{split}
\end{equation}
In analogy to the discussion in 
section \ref{AboutTheCharacteristicEquation} we have to require
$\mathcal{H}{}_R \neq 0$ for a reasonable law of light propagation.
With the same argument as in section \ref{M0Fall} we find that in 
the case $\hat M=0$ at least one of the optical metrics necessarily 
violates the causality condition. In the case $\hat M\neq0$, the
optical metrics (\ref{eq:optmet}) can be rewritten in terms of the 
Hamiltonian as
\begin{equation}
a{}_{\pm}^{ik}=\eta^{ik}+\hat\sigma_{\pm} \tilde{H}{}^{im}\tilde{H}{}^k_{\phantom km}
\quad\text{with}\quad
\hat{\sigma}{}_{\pm}=
\frac{\hat N}{2\hat M}\pm\sqrt{\frac{\hat N^2}{4\hat M^2}-\frac{\hat P}{\hat M}}\,.
\end{equation}
Thereupon, the causality conditions (\ref{ok2}) for a background field with
$\mathcal{H}{}_R \neq 0$ and $\hat{M} \neq 0$ require
\begin{equation}\label{HamOk2}
\hat{\sigma}_A\leq0 \, , \quad \quad
1+\hat \sigma_AR-\hat{\sigma}^2_AS^2>0 \, .
\end{equation}

As for the standard Maxwell vacuum theory $(F_{ab}=H_{ab})$ we have $\mathcal{H} _R = 1/2$,
according to (\ref{eq:constH}), we write down the analogues of the causality conditions 
(\ref{eq:causcon7}), (\ref{eq:causcon8}) and  (\ref{eq:causcon9}) for the case that 
$\mathcal{H} _R >0$. Applying the replacement rules yields
\begin{gather}
\label{eq:causconH1}
\mathcal{H}_{RR} \mathcal{H}_{SS}-\mathcal{H}_{RS}^2\geq0\,,
\\
\label{eq:HRRHSS}
\mathcal{H}{}_{RR} \le 0 \, , \quad \mathcal{H}{}_{SS} \le 0 \, ,\\
\label{eq:causconH3}
\mathcal{H}_R >
-R \mathcal{H}_{RR}+
\dfrac{R}{4} \mathcal{H}_{SS}-2 S \mathcal{H}_{RS}
-\Big(\mathcal{H}_{RR}+ \dfrac{1}{4} \mathcal{H}_{SS}\Big)
\sqrt{R^2+4 S^2} \, ,
\end{gather}
In analogy to the Lagrangian case, the causality conditions for theories with 
$\mathcal{H}{}_R < 0$ follow from the ones for theories with $\mathcal{H}{}_R 
> 0$ by replacing $\mathcal{H}$ with $- \mathcal{H}$ everywhere.

If the Lagrangian is independent of $G$, the Hamiltonian is independent of $S$, i.e., 
$\mathcal{L}(F)$-theories correspond to $\mathcal{H}(R)$-theories. For an 
$\mathcal{H}(R)$-theory with $\mathcal{H}_R>0$ the causality conditions simplify
to
\begin{gather}
\label{eq:causconHR1}
\mathcal{H}{}_{RR} \le 0 \, ,
\\
\label{eq:causconHR2}
\mathcal{H}_R + R \mathcal{H}_{RR} >
- \mathcal{H}_{RR} \sqrt{R^2+4 S^2} \, ,
\end{gather}
In analogy to the Lagrangian case we see that only in the Maxwell case $\mathcal{H}_{RR} =0$
can these conditions hold for arbitrarily large $S$.
  
Here we have chosen the Lagrangian formulation as the starting point and derived the 
corresponding Hamiltonian formulation under the assumption that the constitutive
law (\ref{eq:const}) can be solved for the field strength. Conversely, one could start out from a 
Hamiltonian formulation and then perform the passage to the Lagrangian formulation
provided that the constitutive law (\ref{eq:constH}) can be solved for the excitation. 
We emphasize that there are some examples of interest where the theory 
is formulated in terms of a Hamiltonian and where the constitutive law 
(\ref{eq:constH}) does not uniquely determine the excitation in terms of the field strength. 
In those cases the causality conditions (\ref{eq:causconH1}) to (\ref{eq:causconH3}) are
valid all the same because they can be derived directly from the field equations in terms of the Hamiltonian.  An example of this kind will be treated in section \ref{subsec:regular} below.

\section{Examples}\label{sec:examples}

In this section we briefly investigate the causality conditions  for some 
specific theories of the Pleba{\' n}ski class.

\subsection{Standard Maxwell vacuum theory}\label{subsec:Max}
The standard Maxwell vacuum theory is an $L(F)$-theory with the 
Lagrangian
\begin{equation}
\mathcal{L}=-\frac{F}{2} \,,
\end{equation}
hence
\begin{equation}\label{eq:MaxLF}
\mathcal{L}_F=-1/2\,;\quad \mathcal{L}_{FF}=0\,
\end{equation}
All values of $F$ and $G$ are allowed and the constitutive law reads
\begin{equation}\label{eq:constMax}
H^{mn}=F^{mn}\,.
\end{equation}
From (\ref{eq:MaxLF}) we read that we have indeed $\mathcal{L}_F <0$ and that
the causality conditions (\ref{eq:causconF1}) and (\ref{eq:causconF2}) 
are satisfied in a trivial way for all $F$ and $G$. As in
the standard Maxwell vacuum theory the optical metrics coincide 
with the space-time metric, it is clear from the outset that the 
causality conditions have to hold.

\subsection{Born-Infeld theory}\label{subsec:BI}
The Born-Infeld theory~\cite{BornInfeld1934} derives from the Lagrangian
\begin{equation}\label{BornInfeldLag}
\mathcal{L}(F,G)=-b_0^2\sqrt{1+\frac{F}{b_0^2}-\frac{G^2}{b_0^4}}+b_0^2
\end{equation}
where $b_0$ is a hypothetical constant of Nature with the dimension of 
a field strength. For $b_0 \to \infty$ the Born-Infeld theory approaches the
standard Maxwell vacuum theory. 

From the Lagrangian we find
\begin{equation}\label{eq:LFBI}
\mathcal{L} _F=  
\dfrac{-1}{2 \sqrt{1+ \dfrac{F}{b_0^2}-\dfrac{G^2}{b_0^4}}} \, ,
\end{equation}
\begin{equation}\label{eq:LGBI}
\mathcal{L} _{G} = 
\dfrac{G}{ b_0^2 \sqrt{1+ \dfrac{F}{b_0^2}-\dfrac{G^2}{b_0^4}}} \, ,
\end{equation}
so the constitutive law (\ref{eq:const}) reads
\begin{equation}\label{BIFeldstaerke1}
H^{ab}=\frac{F^{ab} + \frac{G}{b_0^2} \tilde F{}^{ab}}{\sqrt{1+\frac{F}{b_0^2}-\frac{G^2}{b_0^4}}}
\, .
\end{equation}
As the excitation has to be real and finite, the allowed values of $F$ and 
$G$ are 
restricted by
\begin{equation}\label{eq:BI1}
1+\frac{F}{b_0^2}-\frac{G^2}{b_0^4}>0
\end{equation}
which requires, in particular,
\begin{equation}\label{eq:BI2}
1+\frac{F}{b_0^2}>0 \, .
\end{equation}

To demonstrate causality we could check if the causality conditions (\ref{eq:causcon7}),
(\ref{eq:causcon8}) and (\ref{eq:causcon9}) are satisfied. However, in this case we prefer 
to directly verify that the conditions (\ref{eq:sigmain}) together with 
$\mathcal L_F < 0$ and $M > 0$ are satisfied for all allowed 
background fields. We read from (\ref{eq:LFBI}) that $\mathcal{L} _F < 0$ is obviously 
true for all allowed field values. 
By calculating the second derivatives of the Lagrangian we find that
\begin{equation}\label{eq:MBI}
M =  \dfrac{{\Big( 1 + \dfrac{F}{b_0^2} \Big)} ^2}{ 4 \Big( 1+ \dfrac{F}{b_0^2}-\dfrac{G^2}{b_0^4}\Big) ^2} \, ,
\end{equation}
so the condition $M>0$ is satisfied for all allowed $F$ and $G$. 
To verify the inequalities  (\ref{eq:sigmain}) we note that the two optical metrics 
(\ref{eq:optmet}) coincide and that they satisfy 
\begin{equation}\label{eq:sigmaBI}
\sigma _+ = \sigma _- =  \dfrac{-1}{b_0^2+F}
\end{equation}
and 
\begin{equation}\label{eq:detBI}
1+ \sigma _{\pm} F - \sigma _{\pm} ^2 G^2 = 
\dfrac{1 + \dfrac{F}{b_0^2} - \dfrac{G^2}{b_0^4}}{\Big( 1 + \dfrac{F}{b_0^2} \Big)^2} \, .
\end{equation}
This demonstrates that the causality conditions (\ref{eq:sigmain})  
are satisfied precisely for the allowed field values.

\subsection{Born theory}\label{subsec:Born}
Before introducing the Born-Infeld theory, Born~\cite{MB} had suggested
the Lagrangian 
\begin{equation}
\mathcal{L}(F)=-b_0^2\sqrt{1+\frac{F}{b_0^2}}+b_0^2 \, .
\end{equation}
This is an $\mathcal{L}(F)$-theory with
\begin{equation}\label{eq:LFB}
\mathcal{L} _F=  \dfrac{-1}{2  \sqrt{1+ \dfrac{F}{b_0^2}}}
\end{equation}
\begin{equation}\label{eq:LFFB}
\mathcal{L} _{FF} =  \dfrac{1}{ 4b_0^2 {\sqrt{1+ \dfrac{F}{b_0^2}}\,}^3} \, ,
\end{equation}
which leads to the constitutive law
\begin{equation}\label{eq:constBorn}
H^{ab}=\frac{F^{ab}}{\sqrt{1+\frac{F}{b_0^2}}}\,.
\end{equation}
The allowed values of $F$ are restricted by
\begin{equation}\label{eq:restBorn}
1+\frac{F}{b_0^2}>0 
\end{equation}
while there is no restriction on the values of $G$. 

From (\ref{eq:LFB}) and (\ref{eq:LFFB}) we read that
$\mathcal{L}_F < 0$ and that the first causality condition (\ref{eq:causconF1})
is satisfied for all allowed field values. However, the second 
causality condition (\ref{eq:causconF2}) is violated if $G^2 \geq b_0^2 (b_0^2+F)$. 
This exemplifies our general result 
that, for any $\mathcal{L}(F)$-theory different from the standard 
Maxwell vacuum theory, the causality condition is violated for 
background fields with sufficiently big $G$.

\subsection{Heisenberg-Euler theory}
The Heisenberg-Euler theory~\cite{HeisenbergEuler1936} incorporates some effects 
from quantum electrodynamics into an effective classical theory. We will discuss the approximated Lagrangian up two second order following the notation of Dunne~\cite{Dunne}
\begin{equation}\label{eq:HEL}
\mathcal{L}=E_0^2\left\{-\frac12\,\frac{F}{E_0^2}+
\Lambda\left(\frac{F^2}{E^4_0}+7\,\frac{G^2}{E^4_0}\right)\right\}
\end{equation}
where
\begin{equation}
\begin{gathered}
\Lambda=\frac{\hbar c}{90\pi e^2} = 0.7363 
\\
E_0=\frac{m^2 c^4}{e^3} = 
6.048 \times 10^{15}\,\frac{\sqrt{\mathrm g}}{\sqrt{\mathrm{cm}}\,\mathrm s} \, .
\end{gathered}
\end{equation}
The second-order approximation is justified only if the higher-order terms
are small, so the allowed field values are restricted by
\begin{equation}\label{eq:HErest}
F^2 + 4G^2 \ll E_0^4 \, .
\end{equation}
From the Lagrangian we find 
\begin{equation}
\begin{gathered}
\mathcal{L}_F=-\frac12+\frac{2\Lambda\,F}{E_0^2}\,,\quad\mathcal{L}_G=\frac{14\Lambda\,G}{E_0^2}\,,\\
\mathcal{L}_{FF}=\frac{2\Lambda}{E_0^2}>0\,,\quad\mathcal{L}_{GG}=\frac{14\Lambda}{E_0^2}>0\,,\quad\mathcal{L}_{FG}=0\,.
\end{gathered}
\end{equation}
So the first two causality conditions (\ref{eq:causcon7}) and (\ref{eq:causcon8}) are
obviously satisfied and the third one (\ref{eq:causcon9}) requires
\begin{equation}
E_0^2/\Lambda>F+11\sqrt{F^2+4G^2}\,.
\end{equation}
This is, indeed, satisfied for all allowed field values (\ref{eq:HErest}).

\subsection{A pathological example}\label{subsec:path}
For the sake of argument, we consider the Lagrangian
\begin{equation}\label{ModBornInfeldLag}
\mathcal{L}(F,G)=b_0^2\sqrt{1-\frac{F}{b_0^2}-\frac{G^2}{b_0^4}}-b_0^2
\end{equation}
which results from the Born-Infeld Lagrangian by replacing
$(F,G)$ with $(-F,G)$. In a $(3+1)$ splitting this corresponds to 
interchanging the electric and the magnetic field strength. In 
comparison to the Born-Infeld case we have also changed 
the overall sign of the Lagrangian to have again $\mathcal{L}_F < 0$. 
Just as the Born-Infeld Lagrangian, the Lagrangian 
(\ref{ModBornInfeldLag}) approaches the Maxwell
Lagrangian, $\mathcal{L}(F,G) \to -F/2$, for $b_0 \to \infty$.

From the Lagrangian we find
\begin{equation}\label{eq:LFpath}
\mathcal{L} _F=  
\dfrac{-1}{2 \sqrt{1- \dfrac{F}{b_0^2}-\dfrac{G^2}{b_0^4}}} \, ,
\end{equation}
\begin{equation}\label{eq:LGpath}
\mathcal{L} _{G} = 
\dfrac{-G}{ b_0^2 \sqrt{1- \dfrac{F}{b_0^2}-\dfrac{G^2}{b_0^4}}} \, ,
\end{equation}
so we have indeed $\mathcal{L}_F <0$ and 
the constitutive law (\ref{eq:const}) reads
\begin{equation}\label{constpath}
H^{ab}=
\frac{F^{ab} - \frac{G}{b_0^2} 
\tilde F{}^{ab}}{\sqrt{1-\frac{F}{b_0^2}-\frac{G^2}{b_0^4}}} \, .
\end{equation}
The allowed values of $F$ and $G$ are restricted by
\begin{equation}\label{eq:resttpath}
1-\frac{F}{b_0^2}-\frac{G^2}{b_0^4} > 0 \, .
\end{equation}
The second derivative of the Lagrangian with respect to $F$
is given by 
\begin{equation}\label{eq:LFFpath}
\mathcal{L}_{FF}= \frac{-1}{4 b_0 \left(1-\frac{F}{b_0^2}-\frac{G^2}{b_0^4}\right)^{3/2}} \, ,
\end{equation}
which implies that $\mathcal{L}_{FF} < 0$ for all allowed field 
values. However, according to (\ref{eq:causcon8}) $\mathcal{L}_{FF} \ge 0$ 
is a necessary condition for causality. Therefore, in this case the 
causality condition is violated for \emph{all} background fields.

\subsection{Hamiltonian for a regular black hole}\label{subsec:regular}
Several regular black holes have been found which are solutions to Einstein's field
equation coupled to a nonlinear electrodynamical theory of the Pleba{\'n}ski class. 
Here we consider an example that was found by  Ayon-Beato and 
Gar\-c{\' \i}a~\cite{AyonGarcia1999}. For a discussion of general features of such
regular black-hole solutions we refer to Bronnikov~\cite{Bronnikov}.

The electrodynamical theory is given in terms of a Hamiltonian
\begin{equation}
\mathcal{H}(R)
=\frac{R}{2\,\cosh^2 \big( \sqrt[4]{R/R_0} \big)}
\end{equation}
where $R_0>0$ is a constant. 

The derivatives of the Hamiltonian
with respect to $R$ are
\begin{equation}\label{eq:reg1}
\mathcal{H}{}_R=
\dfrac{
2- \sqrt[4]{R/R_0} \, \mathrm{tanh} \big( \sqrt[4]{R/R_0} \big)
}{
4 \, \cosh^2 \big( \sqrt[4]{R/R_0} \big)
}
\, , 
\end{equation}
\begin{equation}\label{eq:reg2}
\mathcal{H}{}_{RR}= 
\dfrac{
 -4 \sqrt{R/R_0} 
-5 \sqrt[4]{R/R_0} \, \mathrm{sinh} \big( 2 \sqrt[4]{R/R_0} \big)
+ 2 \sqrt{R/R_0} \, 
\mathrm{cosh} \big( 2 \sqrt[4]{R/R_0} \big) \Big)
}{
32 \, R \, \mathrm{cosh} \big( \sqrt[4]{R/R_0} \big)
}
\, .
\end{equation}
We see that the allowed values of $R$ are restricted by
\begin{equation}\label{eq:regR}
R \geq 0
\end{equation}
because for $R<0$ the Hamiltonian is non-real. As in any other
$\mathcal{H} (R)$-theory there is no restriction on the values of $S$. 
From (\ref{eq:reg1}) we read that $\mathcal{H}{}_R\rightarrow 1/2$
for $R\rightarrow 0$, so the theory approaches the Maxwell vacuum
theory in the weak-field limit as was emphasised in~\cite{AyonGarcia1999}. 

It is a characteristic feature of this type of regular black-hole solutions that  
the condition $\mathcal{H}{}_R \neq 0$ does not hold on the entire domain
 of allowed values, i.e., that $\mathcal{H}$ goes through an extremum at some 
value $R = R_m$. In the example at hand, one can calculate from (\ref{eq:reg1}) 
that $\mathcal{H}$ goes through a maximum at $R_m \approx 18.2 \, R_0$. 
One can perform the passage to the Lagrangian formalism if one restricts
either to values $R>R_m$ or to values $R<R_m$. However, on any interval that 
contains the point $R=R_m$ the relation between $R$ and $F$ is not one-to-one 
and the passage to the Lagrangian is not possible, cf. Bronnikov~\cite{Bronnikov}.

For our investigation of causality we restrict to the interval  $0 \le R < R_m$ where
$\mathcal{H}{}_R >0$ and we check if the causality conditions (\ref{eq:causconHR1})
and (\ref{eq:causconHR2}) hold on this interval. The first condition
is indeed true on the entire interval under consideration.
However, the second condition is violated, even for fields with $S=0$ if
$R \gtrsim 0.62 \, R_0$. So here we have an example where we do not
need a big value of $S$ to violate the causality condition.

\section{Conclusions}
If nonlinear vacuum electrodynamics is realised in Nature, we have to
distinguish three light-cones: The light-cone of the space-time metric
and the light-cones of the two optical metrics. Light rays are lightlike
geodesics of the optical metrics whereas, according to the standard
interpretation of (special or general) relativity, the light-cone
of the space-time metric determines the maximal speed of signals.
If this interpretation is accepted, the optical metrics should be 
causal in the sense that their lightlike geodesics are lightlike or
timelike with respect to the space-time metric. It was the purpose 
of this paper to characterise those nonlinear vacuum electrodynamical
theories for which this is true.

We proceeded in two steps. In the first step, in 
sections~\ref{Kausalitaet_MNicht0}, \ref{M0Fall} and 
\ref{Abschnitt5}, we gave criteria that can be easily
checked if the Lagrangian and the background field are specified. These
results are complementary to the work of Goulart and Perez 
Bergliaffa~\cite{Costa} and Abalos et al.~\cite{Abalos}. In particular,
we refer to the latter paper for pictures of how the three light-cones
are related in different situations. However, we have prefered to 
derive our results in the easiest way directly from the characteristic 
equation, without refering to results from the quoted papers. In the 
second step, in sections~\ref{sec:causLag} and \ref{sec:causHam}, we 
worked out conditions \emph{on the Lagrangian}, or \emph{on the Hamiltonian}, 
of the electrodynamical theory that guarantee causality of the optical 
metrics for all allowed background fields. These conditions 
were given as a set of inequalities the derivatives of the Lagrangian, 
or of the Hamiltonian, have to satisfy. For an arbitrary 
$\mathcal{L}(F,G)$-theory, or for an arbitrary 
$\mathcal{H}(R,S)$-theory, it might be desirable to further
investigate these conditions and to characterise in an even more
convenient way all theories that guarantee causality. However, for 
$\mathcal{L}(F)$-theories or $\mathcal{H}(R)$-theories the criteria 
can be easily evaluated. In particular, we have seen that any such
theory other than the standard Maxwell vacuum theory necessarily
violates the causality conditions for some allowed background fields.

We have assumed throughout that the underlying space-time metric is the
Minkowski metric. However, this is no restriction of generality. As
the causality conditions are algebraic conditions on the light-cones, 
they can be applied on a curved space-time manifold to the light-cones 
on each tangent space.

We have said that the optical metrics have to be causal if one 
accepts the hypothesis that the light-cone of the 
space-time metric determines the maximal speed of signals. We should
add that this hypothesis is not totally beyond any doubt. Firstly,
one might argue that (tiny) violations might be possible if 
quantum effects are taken into account, in particular on a curved 
background, see Drummond and Hathrell~\cite{DrummondHathrell1980}. 
However, even in this case our causality
conditions remain valid in the sense that they should be satisfied 
to a very good approximation. Secondly, one might argue that the 
light-cone of the space-time metric does not determine the maximal 
speed of signals if there are strong background fields. 
This is certainly
a possibility, but it is tantamount to making major changes to 
the interpretation of the space-time metric, i.e., to the theory of 
relativity. At present, we see no compelling reason why one
should do this. Therefore, we believe that there are
good reasons to consider only those nonlinear vacuum electrodynamical
theories as physical which give causal optical metrics for all
allowed background fields.

\section*{Acknowledgments}
GS wishes to thank Evangelisches Studienwerk Villigst for supporting 
him with a PhD Stipend during the course of this work. VP is grateful 
to Deut\-sche Forschungsgemeinschaft for financial support under 
Grant No. LA 905/14-1. Moreover, we gratefully acknowledge support 
from the Deutsche Forschungsgemeinschaft within the Research Training 
Group 1620 ``Models of Gravity''. GS also wishes to thank 
H. v. Borzeszkowski for helpful discussions and comments.

\bibliography{optmet}
\end{document}